\documentclass[prl,twocolumn,showpacs]{revtex4}  % for review and submission
\usepackage{graphicx}  % needed for figures
\usepackage{bm}        % for math
\usepackage{amssymb}   % for math

\begin{document}

% the following line is for submission, including submission to the arXiv!!
%\hspace{5.2in} \mbox{Fermilab-Pub-04/xxx-E}

\title{Gamma-ray Laser and Radiation from Collimated Particles\footnote{Sov. Tech. Phys. Lett., 1978, Vol. 4(5),  pp.238-239 } }

%\input author_list.tex       % D0 authors (remove the first 3 lines
                             % of this file prior to submission, they
                             % contain a time stamp for the authorlist)
                             % (includes institutions and visitors)
\author{G.\,V.\, Kovalev \/\thanks}

%%% author(s) - for colontitle (at the top of the page)
%\rauthor{G.\,V.\,Kovalev}

%%% author(s) - for table of contents
%\sodauthor{Kovalev}

%%% author's address(es)
\affiliation{Engineering-Physics Institute, Moscow }

%%% dates of submition & resubmition (if submitted once, second argument is *)
%\dates{13 March 1978}

\date{Mar. 13, 1978}

\begin{abstract}
The motion of channeled particles is accompanied by the photon emission. This feature can be used for the stimulated  generation of high energy photons, but the required density of channeled particles must be very high.
\end{abstract}

%%% PACS numbers
\pacs{ 61.80.Mk}
\maketitle

%\section{\label{sec:level1}First-level heading}
% sections are not used for PRL papers
The radiation emitted by a fast charged particle collimated by a crystal shows the following spectrum [1,2]
\begin{eqnarray}
\frac{d I}{d \omega}=e^2  \rho^2_{1 2}  \Delta \, \epsilon^{2}_{1 2} \omega [1- 2\frac{\omega}{\omega_{max}}+2\left(\frac{\omega}{\omega_{max}}\right)^2], \nonumber \\
\omega_{min} < \omega  < \omega_{max}, \nonumber \\
\omega_{max}=2  \Delta \, \epsilon^{2}_{1 2}  \gamma^2, \,\, \omega_{min} =\frac{ \Delta \, \epsilon^{2}_{1 2} }{2},
\label{r1}
\end{eqnarray}
where $\gamma = (1- v^2)^{-1/2}$ is the relativistic factor $(\hbar = c = 1)$;
$e  \rho_{1 2}$ is the dipole moment of the transition from level 2
to level 1, and $ \Delta \, \epsilon_{1 2} $ is the spacing between the levels for
bounded transverse motion. With planar collimation of
positrons (or electrons) in a parabolic potential this spacing is $ \Delta \, \epsilon_{1 2} \simeq 0.1 \gamma^{1/2}$ eV ($ \Delta \, \epsilon_{1 2} \simeq 10 \gamma^{-1/2}$ eV  for electrons); in a box potential with infinitely high walls the spacing is $ \Delta \, \epsilon_{1 2} \simeq 0.1 \gamma^{-1}$ eV ($ \Delta \, \epsilon_{1 2} \simeq 10 \gamma^{-1}$ eV) [3].

The opposite situation holds in axial collimation of electrons. In this case the spacing is $ \Delta \, \epsilon_{1 2} \simeq 10 \gamma$  eV  [4]  (the spacing between transverse levels increases with in creasing particle energy), and the maximum spectral intensity increases rapidly according to the dipole approximation ($\rho_{1 2} \gamma \Delta \, \epsilon_{1 2} << 1$), since $(\frac{d I}{d \omega})_{\omega=\omega_{max}}  \sim  \gamma^4$  from (1).

This feature can be exploited to get induced emission of photons with an energy $\sim 10$ eV. The directional variation
of the radiation frequency in collimation is governed by the Doppler effect[5],  $ \omega=\omega_0 \gamma^{-1} (1-v \cos \theta)^{-1}$,   where $\omega_0 = \gamma \Delta \, \epsilon_{1 2}$.  In a laser without mirrors, in which the direction of the stimulated emission is governed by the minimum damping of the photon or the geometry of the working volume (for low damping), in the direction perpendicular to the longitudinal velocity of the particle (where the damping is low; see the discussion below)  we have $\omega =\Delta \, \epsilon_{1 2}  \gamma$ for electrons but $\omega =\Delta \, \epsilon_{1 2}  \gamma^{-1/2} $ for positrons. 
At a higher energy of the electron beam it is possible, in principle, to achieve stimulated emission well into the $UV$ region and even in the x-ray region. The beam of collimated  particles must be extremely intense, as follows from the estimates below. The required conditions can probably be met for electrons.

We seek the threshold current density $j_{th}$  required
for laser action. The threshold population inversion of
the particles should be [6]
\begin{eqnarray}
\Delta N_{th}= (N_2 - N_1)_{th}=\frac{\tau_{12}}{\tau_{c}} \rho , 
\label{r2}
\end{eqnarray}
where $\tau_{12}$ is the time required for a spontaneous transition of a particle from state 2 to state 1, $\tau_{c}$  is the photon lifetime in the active region, and $\rho $ is the number of different oscillation modes within the photon spectral line.

On the other hand, the population inversion of the particles in the channel is
\begin{eqnarray}
\Delta N= j \frac{S l}{v} (\alpha_{2}^{2}-\alpha_{1}^{2}), 
\label{r3}
\end{eqnarray}
where $j$ is the incident current density, $S$ is the beam 
cross section, $v$ is the particle velocity, $\alpha_{1,2}$ is the condition for matching the plane wave outside the crystal to the wave functions of the particle inside the crystal, $l = 
min(L, l_{coh})$, $L$ is the crystal thickness, and $l_{coh}$ is the 
relaxation length for equilibrium between levels due to 
inelastic and thermal scattering. From (2) and (3) we find
\begin{eqnarray}
j_{th} =\frac{\tau_{12}}{\tau_c} \rho \frac{v}{l S  (\alpha_{2}^{2}-\alpha_{1}^{2})}.
\label{r4}
\end{eqnarray}
The threshold population inversion $\Delta N_{th}$ is invariant 
against a change in the coordinate system, so we can calculate this quantity in the moving system, in which the 
radiation frequency is isotropic if recoil is ignored since it is governed by the radiation of a "one-dimensional" or 
"two-dimensional" atom. For a Lorentzian line shape, $\rho = 8 \pi^2 \omega_0^{3} \frac{\Delta \omega_0}{\omega_0} V'$.   The emission probability in this system is $w_{12} =\frac{4}{3} \omega_0^{3} e^{2} {\rho'}_{12}^{2} $,  so that
\begin{eqnarray}
 \Delta N_{th}= 6 \pi^2 \frac{V'}{e^2 {\rho'}_{12}^2 } \frac{\Delta \omega_0}{\tau'_{c} \omega_0}=6 \pi^2 \frac{V}{e^2 \rho_{12}^2 } \frac{\Delta \omega_0}{\tau_{c} \omega_0},
\label{r5}
\end{eqnarray}
where $V'$ and  $\tau'$ are the volume of the active region and 
the photon lifetime in the moving coordinate system.  The 
line width is governed primarily by the energy spread of 
the beam; that is, $ \frac{\Delta \omega_0}{\omega_0} \simeq \frac{\Delta \gamma}{\gamma} $. Since $l <<S^{1/2}$ for ordinary beams, only the transverse modes have the 
minimum damping $\tau_{c}^{-1}$.  With  $\tau_{c} \sim  S^{1/ 2} = 10^{-11}$ sec,  $\rho_{12} = 
10^{-8} $cm, $ (\alpha_{2}^{2}-\alpha_{1}^{2}) \simeq 10^{-3}$,  and  $E_{e-} = 100$ MeV, we find the 
result  $j_{th} = 10^{25}$ particles$/(cm^2 \cdot sec)$.

The value of  $j_{th}$ can be reduced, however, by reducing the energy spread of the beam. Furthermore, we 
can expect an important decrease in  $j_{th}$  in multilayer 
structures at high beam energies, in which the dipole approximation cannot be used ($\rho_{12} \gamma \Delta \epsilon_{12} \geq 1$. It can be shown 
that in this case the radiation emitted in the transition from one medium to another with "shaking" has the spectrum 
\begin{eqnarray}
\frac{d I}{d \omega}= \left \{\begin{array}{ll} I_0 \frac{\omega}{\omega_{max}} (1-\frac{\omega}{\omega_{max}} ) [1-4\frac{\omega}{\omega_{max}}+4\frac{\omega}{\omega_{max}}], \nonumber \\
\omega_{min} < \omega  < \omega_{max}; \nonumber \\
0, \,\,\,  \omega> \omega_{max}, \,  \,\, \omega < \omega_{min}, 
\end{array} \right. 
\label{r6}
\end{eqnarray}
where $I_0 \simeq e \frac{\Delta \, \epsilon_{1 2}}{m -\gamma \Delta \, \epsilon_{1 2}} \left| \frac{Z_2-Z_1}{Z_2}\right|$
and $Z$ is the charge of the atoms of the medium. Assuming an energy $E_{e-} = 10^{11}$ eV,
we find $\tau_{12} \sim 10^{-17}$ sec from (6) and thus $j_{th} = 10^{19}$ particles$/(cm^2 \cdot sec)$.

The author is grateful to N. P. Kalashnikov for  fruitful  discussion.

%\bibliographystyle{jetpl}
%\bibliography{../../Focusing_and_Channeling_in_Crystals/chan02}

\end{document}